\documentclass{article}
\usepackage{amsmath}
\usepackage{amssymb}

\usepackage{graphicx}

\usepackage{cite}

\usepackage{color}

\usepackage[labelfont=bf,labelsep=period,justification=raggedright]{caption}

\makeatletter
\renewcommand{\@biblabel}[1]{\quad#1.}
\makeatother

\begin{document}

\begin{flushleft}
{\Large
\textbf{The topology of a discussion: the \#occupy case}
}
\\
Floriana Gargiulo$^{1}$
Jacopo Bindi$^{2}$, 
Andrea Apolloni$^{3,\ast}$, 
\\
\bf{1}  NaxYs, Unamur, Namur, Belgium
\\
\bf{2}  DISAT  and Center for computational Sciences, Politecnico di Torino,  Torino, Italy
\\
\bf{3}  Department of Infectious Disease Epidemiology, London School of Hygiene and Tropical Medicine, London, United Kingdom
\\
$\ast$ E-mail: andrea.apolloni@lshtm.ac.uk
\end{flushleft}

\begin{abstract}
{We analyse a large sample of the Twitter activity developed around the social movement 'Occupy Wall Street' to study the complex interactions between the human communication activity and the semantic content of a discussion. \\
We use a network  approach based on the analysis of the bipartite graph @Users-\#Hashtags and of its projections: the 'semantic network', whose nodes are hashtags, and the 'users interest network', whose nodes are  users.  In the first instance, we find out that discussion topics (\#hashtags) present a high heterogeneity, with the distinct role  of the communication hubs where most the 'opinion traffic' passes through. In the second case,  the self-organization process of  users activity leads to the emergence of two classes of communicators: the 'professionals' and the 'amateurs'. 
Moreover the network presents  a strong community structure, based on the differentiation of the semantic topics, and a high level of structural robustness  when  a certain set of topics are censored and/or accounts are removed. \\
Analysing the characteristics the @Users-\#Hashtags network we can distinguish three phases of the discussion about the movement. Each phase corresponds  to specific moment of the movement: from declaration of intent, organisation and development  and the final phase of political reactions. Each phase is characterised by the presence of specific \#hashtags in the discussion.}
{Twitter, Network analysis}
\end{abstract}

\section{Introduction}

The number of Twitter users  has experienced a huge increase in the last  years, and today Twitter  counts almost half billion users worldwide (http://semiocast.com/).
Twitter is a micro-blogging  platform , where users can share 140 characters messages, called tweets. Users are identified by the @ symbol proceeding the names, while topics of the tweets are identified by the symbol \# preceding it. These semantic elements are called hashtags and allow the direct search of the tweets concerning a certain topic. Since tweets are in principle visible by all the users, Twitter is meant  more  as an information sharing service than a social network creator.
To increase the visibility of a tweet, users can re-tweet other tweets or  introduce hashtags in the tweets and engage in conversations with other users, who couldn't been reached.
Almost  340 millions of tweets (blog.Twitter.co) are sent every day that can be  collected using freely available  application programming interface(API). In this way, Twitter represents a gold mine of data in a variety of fields including analysis of social network communities \cite{Bryden:2013wv},  Influenza detection\cite{MarcelSalathe:2012ez} \cite{Achrekar:2011dv}, political topics \cite{Tumasjan:2010ue}, sentiments about childhood vaccination\cite{Brooks:uo}  and  predictability of social events\cite{Ciulla:2012wn}. The number of study, based on the analysis  of Twitter, is so big, that  some scientists also have point the importance of ethical guidelines for this type of research, as in the case of animal or human involvement\cite{Rivers:2014kh}.
 Twitter has been recognised to play an important role for the organisation and communication of the recent civil uprisings in North Africa (the Arab Spring)\cite{Howard:2011vf}, the London riot in 2011 \cite{BeguerisseDiaz:2014tl}, the Hong Kong Occupation and the Occupy Wall Street\cite{Conover:2013kt}\cite{ Conover:2013du}. Previous works on the relation between these movements and social media have focused on describing how Twitter has helped  the spread of information, the  characteristics of the users and their geographical distribution. 
To our knowledge, there are few studies that  have focused on how the participants activity and the political discussion has evolved during the period of the movements. 
In this work we use Twitter data  to study  how the political discussion has evolved before, during and after  the Occupy Wall street movement, identifying how the users' social network  and discussion topics have changed during the time. We use social network tools to identify the role of particular users  and topics in the spread of information and in the evolution of the information content.
 The work is organised in this  way: in section \ref{sec:Data} we provide some information about the social movement and  how collected data have been used to construct  the two networks ( one among  users and the other among hashtags) that will use for the analysis; in section \ref{sec:Results} we present the results of the network analysis, the classification of users and the evolution of the discussion; in section \ref{sec:Semantic} we present results on the analysis of the semantic community structure; in section \ref{sec:Temp} we present the results on the evolution of the communities and the network during the movement and finally in section \ref{sec:Discussion}  we draw some conclusions.
 \section{Data and network setup} \label{sec:Data} 
\begin{figure}[htpb]
\centerline{\includegraphics[width=0.5\textwidth]{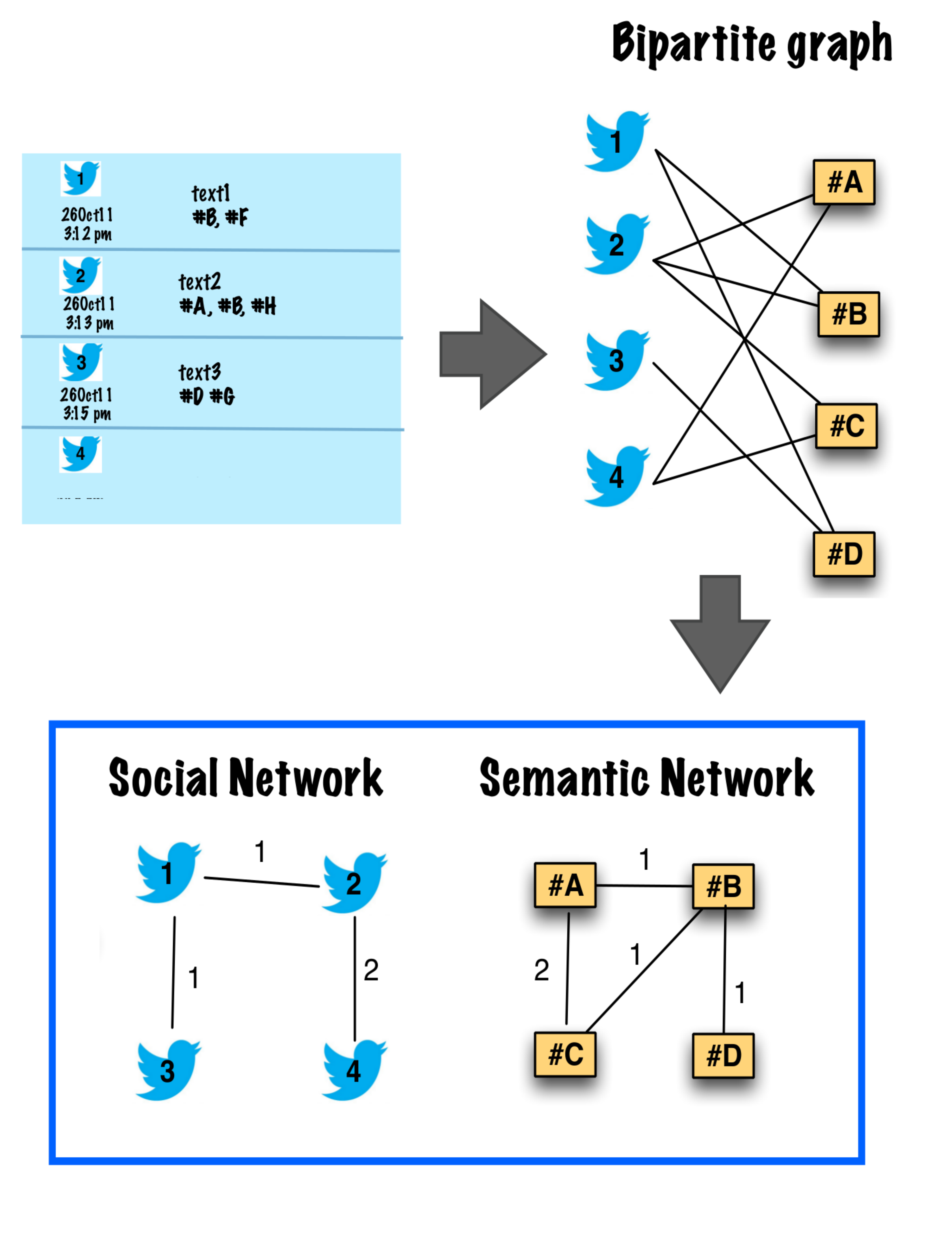}}
\caption{Construction of the bipartite graph from the raw data and projection on the semantic and interest mono-partite graphs. \label{fig1}}
\end{figure}

  On the 13th July 2011,the counter-cultural magazine Adbusters publicised a call to occupy wall street financial district on the following September 17th (https://www.adbusters.org/). Following the call to action, on that day almost 200000 people  gathered at Zuccotti Park.  This date marked the birth of the 'Occupy Wall Street' movement. Occupy movement organised massive demonstrations till May 2012. In few weeks the public space occupation practice spread over about 80 cities in USA and the name 'Occupy', as well.  Occupy activists used Twitter either to organise operations on the ground either to discuss and share news, ideas and  other important subjects. Hashtag as '\#Occupy', '\#ows', '\#occupyoakland' played a key role in Occupy movement communication and gained large exposure in Twitter global activity, too. 
Based on the analysis of a set of tweets containing the word "Occupy" in the text or the hashtag "\#Occupy", we focus on the relations between the users' behaviours and the semantic content of the discussion. The dataset  includes 180K short messages, collected 26th of October 2011 and 18th of April 2012, written by more than 37K  different users and contains 17K hashtags. 
 To study the interaction between these tow aspects we used method similar to the one used by Roth \cite{Roth:2010ff} for analysing the blogosphere activity. Our raw data have been organised to form a bipartite graph @Users-\#Hashtags (\ref{fig1}) using standard techniques  as in \cite{Eubank:2004ca} \cite{Guillaume:2004vl} \cite{2001PhRvE..64b6118N}: to each @user a list of \#hashtags  has been associated; a link is drawn between users  who have used the same hashtag in a tweet;  a link is created between two hashtags used by a user. This procedure defines the two mono-partite network projections: the 'semantic network' for the \#hashtags and 'interest network' for the @users. The number of @users using the same couple of \#hashtags provides the weight of the link. Similarly, in the @users network, two nodes are connected if they use the same \#hashtag.    
The hashtag  \#Occupy  appears in most of the tweets, giving origin to an almost fully connected network and represent a first level of hierarchy of the network. To elicit  the structures created in the two projected networks due to users interests, we have excluded the \#Occupy hastags in the network construction. 
\section{Network Analysis Results}\label{sec:Results}   
\subsection{The static topological structures}
\subsubsection{The semantic network}
\begin{figure}[htpb]
\centerline{\includegraphics[width=1\textwidth]{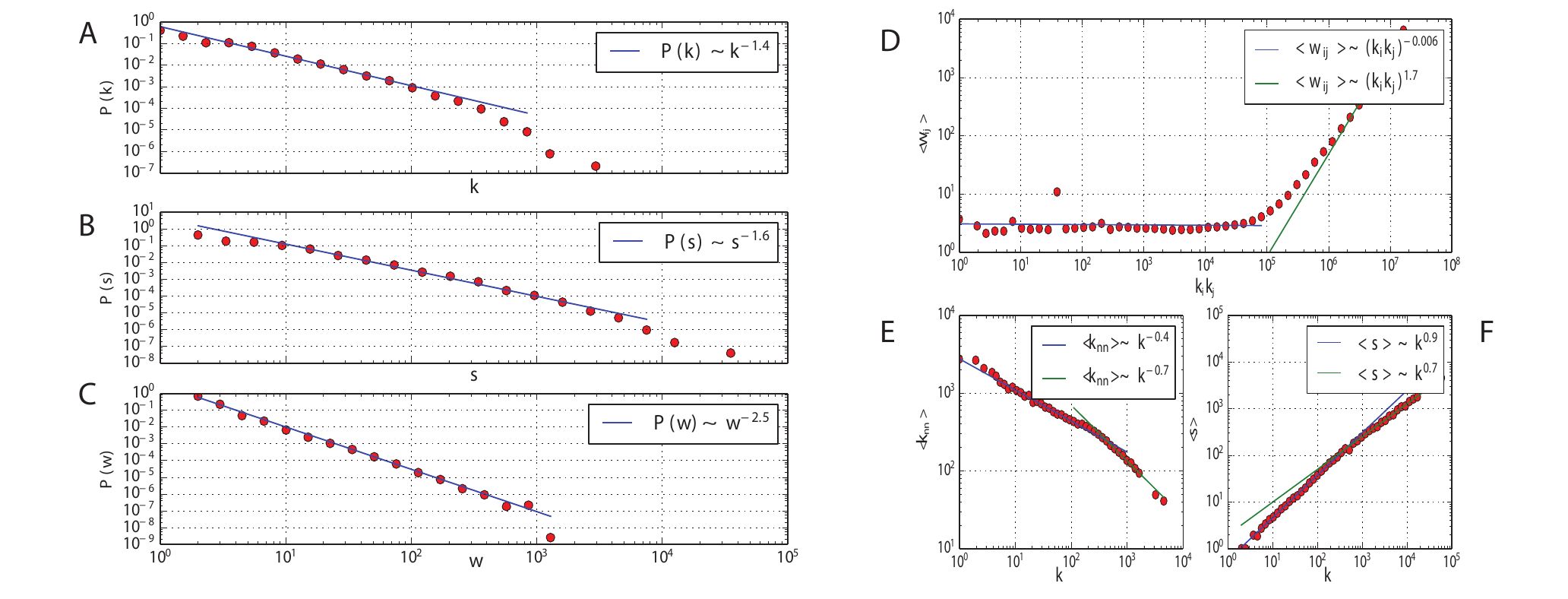}}
\caption{Plot A: Degree distribution of the network. The distribution follows a power law with exponent $\gamma=1.4$. Plot B: Strength distribution of the network. The distribution follows a power law with exponent $\gamma=1.6$. Plot C: Weight distribution of the network. The distribution follows a power law with exponent $\gamma=2.5$. 
Plot D: Average weight values as a function of the product of the degrees at the ends of the edges. In random networks the aspected relation is of the type: $\langle w_{ij}\rangle\sim(k_ik_j)^{0.5}$ Plot E: Degree correlation of the network. The average neighbours' degree decreases sub-linearly with the degree. The decrease is faster for hubs. Plot F: Average strength as a function of the degree. \label{fig2}}
\end{figure}

The semantic network is formed by 5206 hashtags, connected by 95543 links.
This network presents a scale free network structure.
In fact, as shown in Figure \ref{fig2}A,B,C the nodes' degree, the nodes' strength and links' weight distributions, all present an evident power law behaviours: few semantic hubs are present, through which most of the  ''opinion traffic" is channeled through.
Because of this, a clear hierarchy in the semantic network is defined: all the heaviest weighted links  connect network hubs that in turn have strong ties  only with other hubs. A numerical proof of the prevalence of strong ties among hubs is given in Fig. \ref{fig2}D showing that the weight is independent from the product of the nodes degrees for low degrees nodes, whilst grows extremely fast with the product of the nodes degrees for hubs.\\
Regarding  the network's mixing properties  (Fig. \ref{fig2}E), we  notice that the average degree of the neighbours, $k_i^{nn}=(\sum_{j\in\mathcal(v)_i}k_j)/k_i$,
decrease with $k$, denoting a preference for low degree nodes to be connected to high degree ones. Further analysis suggests that hubs' local topology is different from the low degree nodes. For example, network hubs exhibit a stronger preference for connecting to low degree nodes. The strength of the nodes, i.e. the sum of the links weights originating at the node, scales sub-lineraly with the degree (Fig. \ref{fig2}F). This effect is mostly due to the weights heterogeneity and to the fact that for most of the links $w=1$ (i.e. only a user is putting these two hashtags in the same tweet). Since strong ties are equally shared among the hubs,  the low weights of the connections between hubs and low degree nodes, gives origins to the stronger sub-linearity of the hubs behaviour. 
\subsubsection{The users' interest network}
\begin{figure}[htpb]
\centerline{\includegraphics[width=1\textwidth]{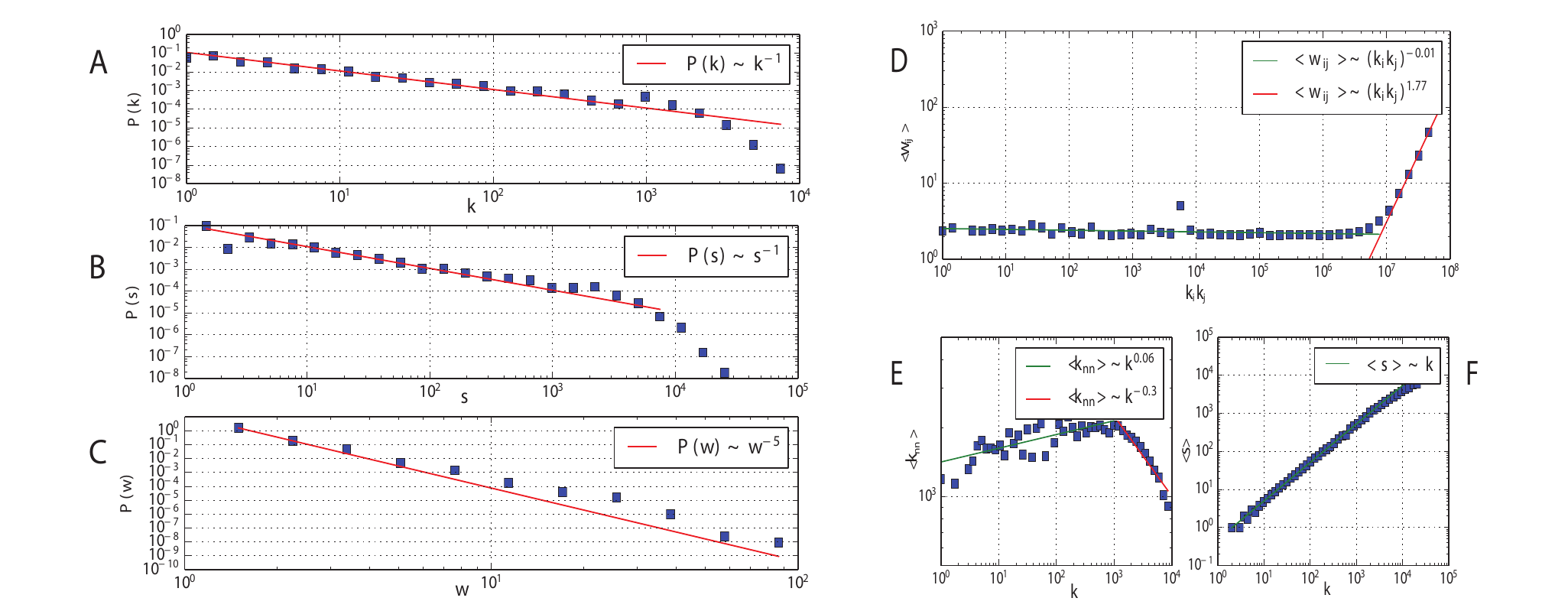}}
\caption{Plot A: Degree distribution of the network. The distribution follows a power law with exponent $\gamma=-1$. Plot B: Strength distribution of the network. The distribution follows a power law with exponent $\gamma=-1$. Plot C: Weight distribution of the network. The distribution follows a power law with exponent $\gamma=-5.1$. Plot D: Degree correlation of the network. The average neighbours' degree increases very slowly with the degree for $k<10^3$. For the hubs ($k>10^3$) we observe a sub-linear decrease. Plot E: Average strength as a function of the degree. Plot F: Average weight values as a function of the product of the degrees at the ends of the edges. In random networks the aspected relation is of the type: $\langle w_{ij}\rangle\sim(k_ik_j)^{0.5}$ \label{fig3}}
\end{figure}

The users' interest network is composed by 12485 users connected by 3962037 links.
The users' interest network is also characterised by power law behaviour for  degree and strength distributions with a low slope (exponent $\gamma=1$, Fig. \ref{fig3}A,B). Both the degree and the strength distribution have an evident exponential cutoff at the tail of the distribution.  Also the weights are distributed according to a power law (Fig. \ref{fig3}C), with a very steep slope. This means that weights heterogeneity is not so strong and therefore the weighted structure is not so relevant for understanding the topology.  \\
Following the procedure  in \cite{Ahn:vz} we distinguish two classes of communicators based on their topological properties: the 'professionals' and the 'amateurs'. Professionals communicators, like bloggers, journalist, official media representative and spybots are concentrated in the tail of the distribution. Professionals are largely connected  (high degree) however  the connections are towards less connected nodes. This feature is typical of communication networks, where the structure is meant to maximise the spread of the communication.   On the other hand  (Fig. \ref{fig3}D), 'amateurs'  or occasionally users, have few connections (indicating a limited activity as a followers or followed) and try to interact mostly among themselves. 

\subsection{Network robustness}
An important feature for social movements is to understand if, and under which condition, the use of Twitter as a communication platform, is robust to repression measures. To this end, we analysed the relative size of the percolation cluster (largest connected component) as we remove from the network the most connected nodes (degree targeted attacks). The sequential removal of nodes in the semantic network mimics censorship measures of the counterpart while, in the users' network it represent the accounts' removal.  
We compare the results with the same procedure on two different reshuffled graphs: the first reshuffling method ("\emph{ab--initio} method")   randomises the position of the existing hashtags in all the messages, that is the hashtags are randomly distributed among users. The second reshuffling method ("topological method") consists in the construction of graphs with the same degree sequence using the configuration model \cite{Molloy:1995tw}.  
\begin{figure}[htpb]
\centerline{\includegraphics[width=1\textwidth]{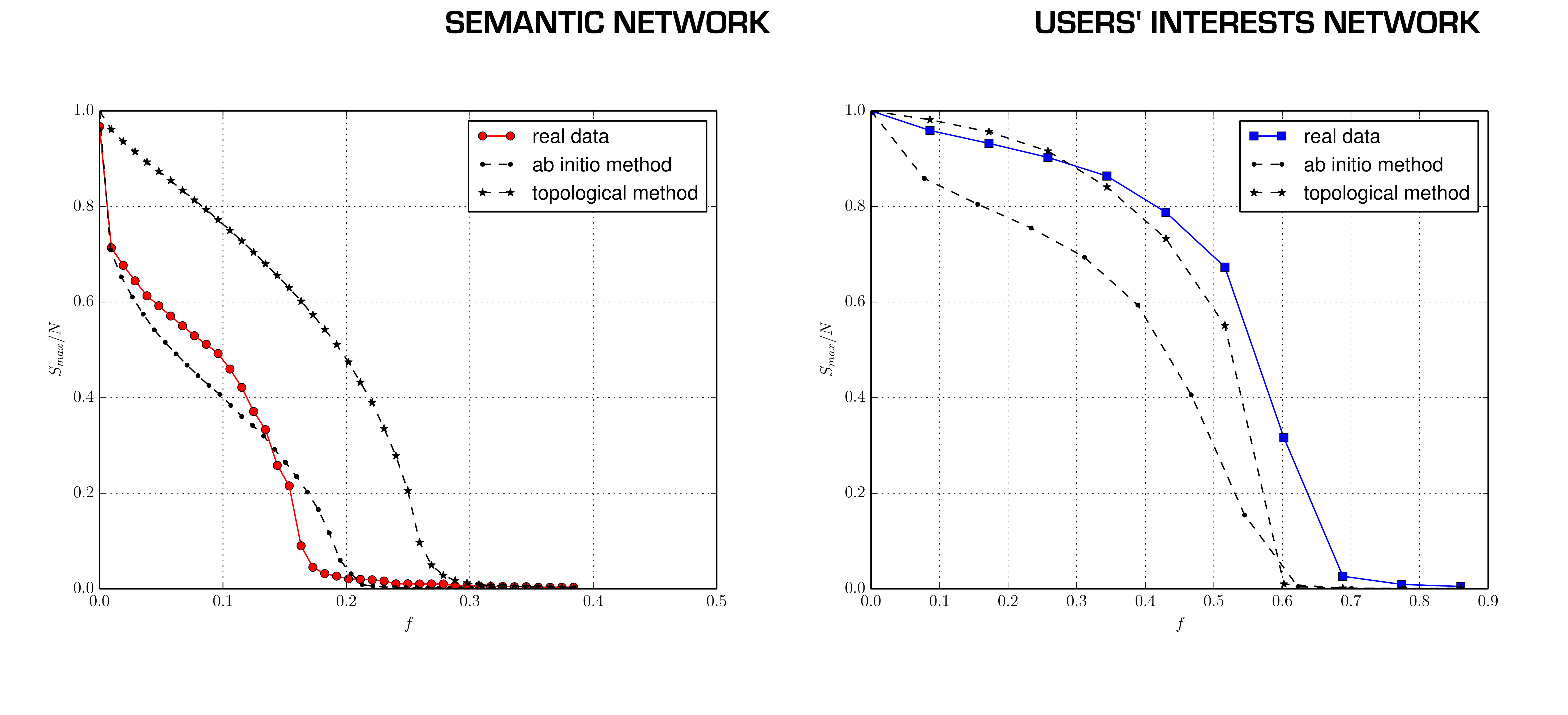}}
\caption{ Robustness of the two mono projections of the network as a function of the removal of nodes (target strategy). On the left semantic network, on the right users interest one. The lines correspond to the  original network (line with coloured dots), the re-shuffled \emph{ab--initio} (dashed line) and the topological method (star-lined)  \label{fig4}}
\end{figure}
The semantic network is less robust than its counterpart after topological reshuffling as seen in Fig.\ref{fig4}A. This is due to the strong hierarchical structure of the network since most of the connections are toward hubs. The topological reshuffling reduce the disassortative behaviour of the network, increasing the robustness.  The \emph{ab--initio} reshuffling for the semantic networks,  increases the number of links between low degree nodes, thus ensuring still connection among them, making the network more able to resist to targeted attacks.  \\
As we can observe in figure \ref{fig4} B the users' network is  more robust to users' deletion than its counterpart after topological reshuffling: that is a large number of users should be censored, before impeding the flow of information. This effect is mainly  due to the high connectivity among amateurs.  These users are the core of the activists deeply involved in the discussions about the movement: they share among them only the very important keywords for the movement filtering out other information. Their strong connectivity is able to guarantee to the network a high level of robustness.  Users' interest graph results to be also much more robust respect to its \emph{ab--initio} reshuffled copies due to the low value of the weights (due to the differentiation of users' semantic preferences): the \emph{ab--initio} reshuffling creates in this case less connections with stronger ties. \\

Twitter is a robust communication tool. The robustness can be improved creating more links among less popular  hashtags (low degree nodes in the semantic network). From a user point of view, this can be achieved, adding in the tweets new hashtags or uncommon ones together with the most used hashtags. In this way, in the case of censorship of most used hashtags, the information can still lows passing through less used ones. 

\section{The semantic clustering}\label{sec:Semantic}
\begin{figure}[htpb]
\centerline{\includegraphics[width=0.7\textwidth]{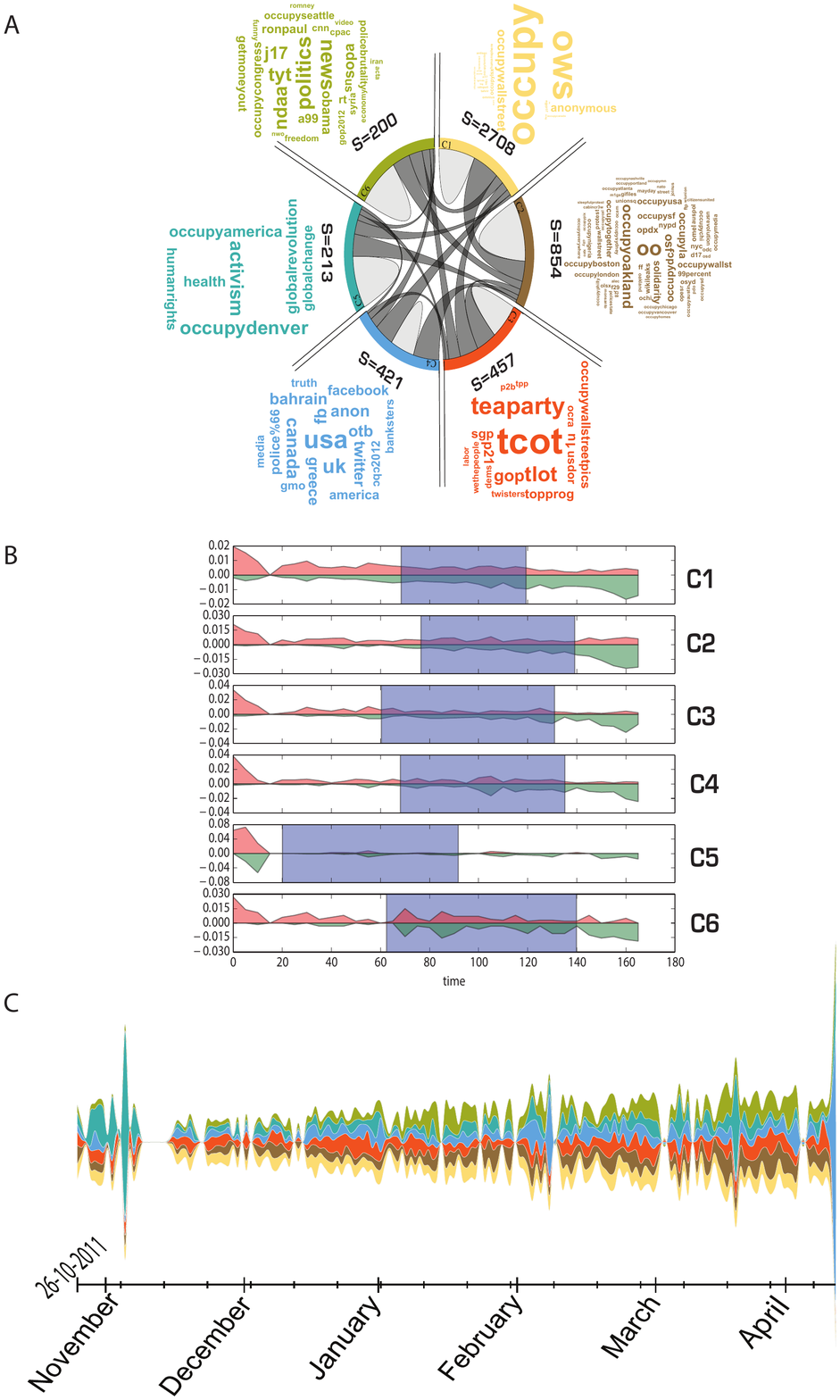}}
\caption{Communities description of the semantic network. The semantic network can be divided in 6 communities.  Plot A: chord diagram representing the interactions among communities and their strengths. Plot B: the permanence time of hashtags in the communities. The red histograms indicate the fraction of  hashtags entering the community at a specific time; the green one the leaving time The shaded area corresponds to the average permanence time Plot C: steam graph of the communities. The  height of the band corresponds to the proportion activity of the community (number of tweets) normalised to the total activity  \label{fig5}}
\end{figure}

To enter deeply into the details of the content of the discussion, we analysed the community structures in the semantic network, using the Louvain community detection method \cite{Blondel:2008vn}. The semantic network is composed by six macroscopic communities with a clear social interpretation (Fig.\ref{fig5}). 
The network hubs are equally distributed among the communities and perform the connections among the different topical areas Fig.\ref{fig5}A. \\
We define the entering time in the discussion (first appearing of the hashtag, $t_i^{MIN}$) and the exiting time (last appearing, $t_i^{MAX}$).
To define the time span of a communities as the average of the permanence times of its members $T_I=(\sum_{i\in I}t_i^{MAX})/N-(\sum_{i\in I}t_i^{MIN})/N$. In general, most of the hashtags are introduced at the very beginning of the discussion and exit at the end. In fact, most of the communities have similar initial and final times, centred around the central day of the data collection. This is true for all the communities except  for  the community C5, that is characterised by the use of very early created  hashtags Fig.\ref{fig5}B. It is important to notice that the first days of the data collection correspond to the phase when the movement was occupying Zuccotti Park (17th September 2011, 15th November 2011) Fig.\ref{fig5}C. The C5 community therefore contains the important themes discussed during the assembles in Zuccotti Park (\#poverty, \#liberty, \#humanrights,...) and defines some of values that have characterised the movement. After the 15th  November the movement continues its discussions mostly on social media platforms, specifying its important topical hashtags. The contents of community C5 are mostly shared with the first  (C1) and the second (C2) communities. The community C1 (the largest one) contains all the important keywords connected to the different typologies of world-spread activism in this kind of social movements (\#ows, \#tharir, \#anonymous, \#revolution). The community C2 contains several local denominations of the Occupy movement (\#oakland, \#cal, \#philly) and significative dates (\#j20,\#j28,\#j29, \#d7 but also \#mayday, \#1m). This is therefore  a community mostly addressed to the organisation of the events. The third community (C3) is mostly focused on discussions around the elections and the institutional politics (\#teaparty, \#tcot,\#tlot). The fourth community  (C4) concerns the internationalisation of the occupy movements (\#eu,\#greece)  and several movement keyword that become international (\#99\%,\#banksters). The last community (C6), for which an high number of tweets were introduced after the eviction of Zuccotti Park, mostly contains discussions about the repression of the movement (\#policebrutality,\#pepperspray).
     
\section{Temporal features}\label{sec:Temp}

\begin{figure}[htpb]
\centerline{\includegraphics[width=0.5\textwidth]{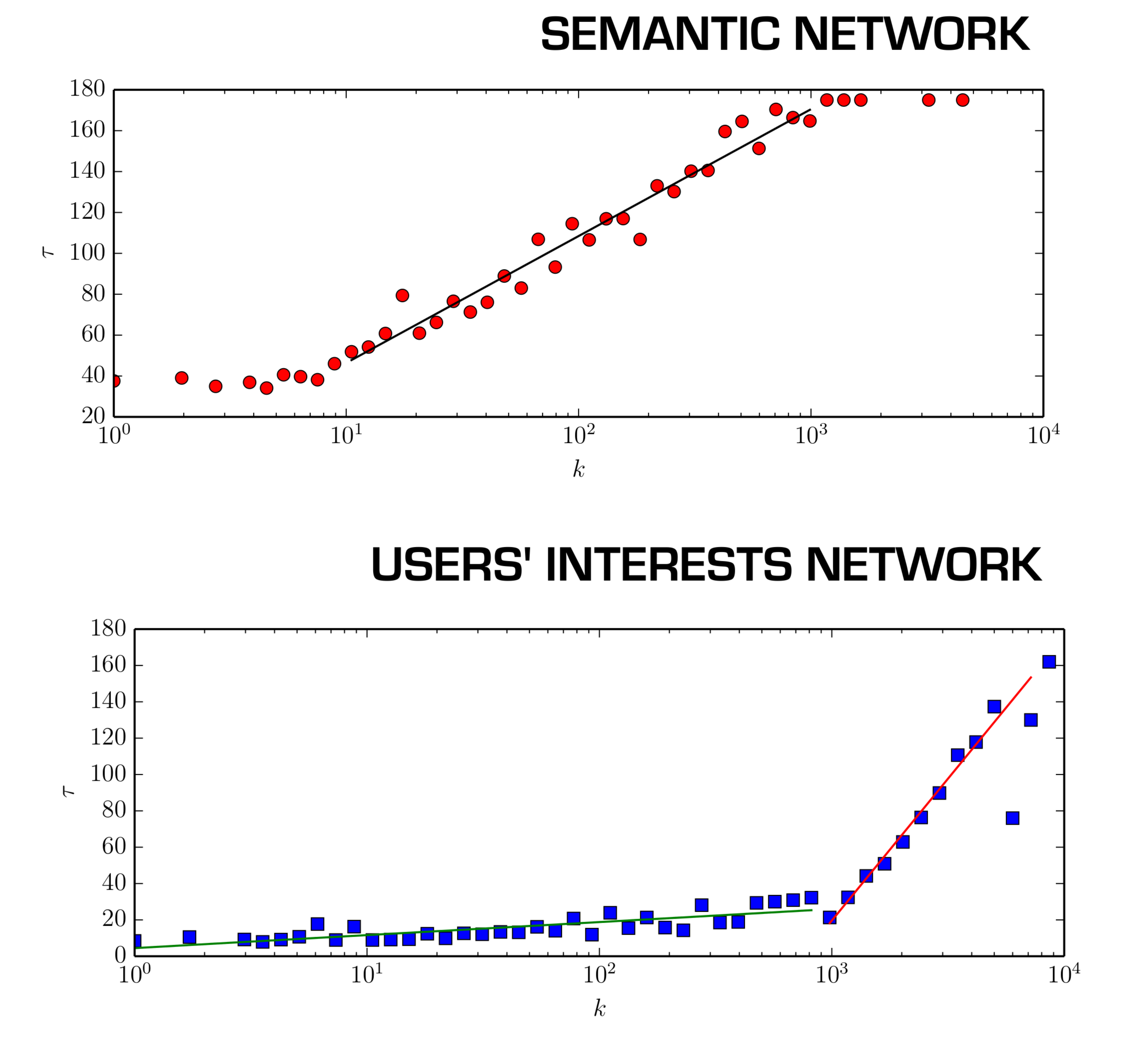}}
\caption{Permanence time as a function of the nodes degree. Top plot: semantic network. Bottom plot: users' interest network\label{fig6}}
\end{figure}
It is trivial to observe that the in the semantic network, hubs have higher permanence (almost  the entire time span of the data collection). At the same time it is obvious that less important nodes just appear for a short period after which they are forgotten (Fig. \ref{fig6}A). On the other hand,  it is interesting to notice that for  the permanence times do not grow linearly with the network degree but logarithmically (Fig. \ref{fig6}B). Whilst the permanence time increase linearly and at the same rate for all the nodes in the semantic network, in the users network, the permanence time for 'amateurs' (slow), almost constant and short, is strongly different from 'professionals' one (fast), that is actually increases as the activity of the user increases.

\begin{figure}[htpb]
\centerline{\includegraphics[width=0.7\textwidth]{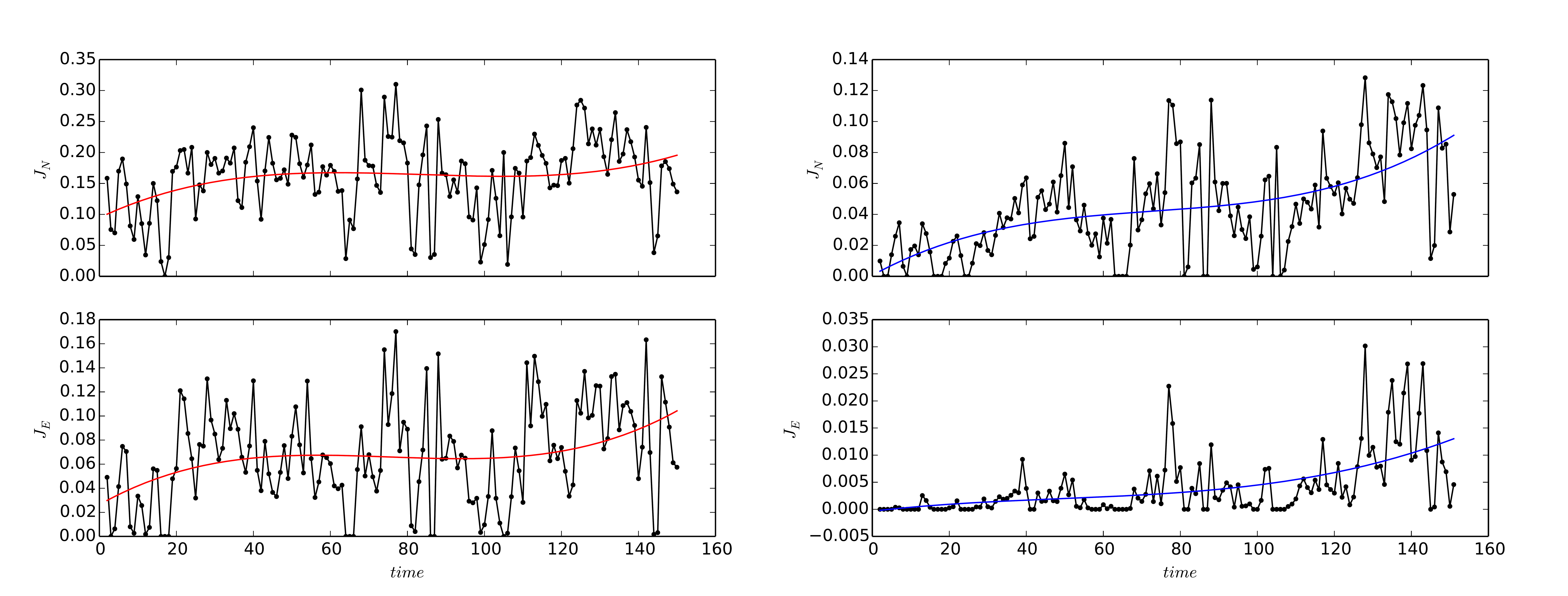}}
\caption{Upper left plot: The black dots and line represent the values of the Jaccard index between the set of nodes of the semantic network at time $t$ and at time $t-1$. The continuous red line is the univariate spline function of the time series. Lower left plot: The black dots and line represent the values of the Jaccard index between the set of edges of the semantic network at time $t$ and at time $t-1$. The continuous red line is the univariate spline function of the time series. Upper right plot: The black dots and line represent the values of the Jaccard index between the set of nodes of the users' interest network at time $t$ and at time $t-1$. The continuous blue line is the univariate spline function of the time series. Lower right plot: The black dots and line represent the values of the Jaccard index between the set of edges of the users' interest network at time $t$ and at time $t-1$. The continuous blue line is the univariate spline function of the time series.\label{fig7}}
\end{figure}

\begin{figure}[htpb]
\centerline{\includegraphics[width=1\textwidth]{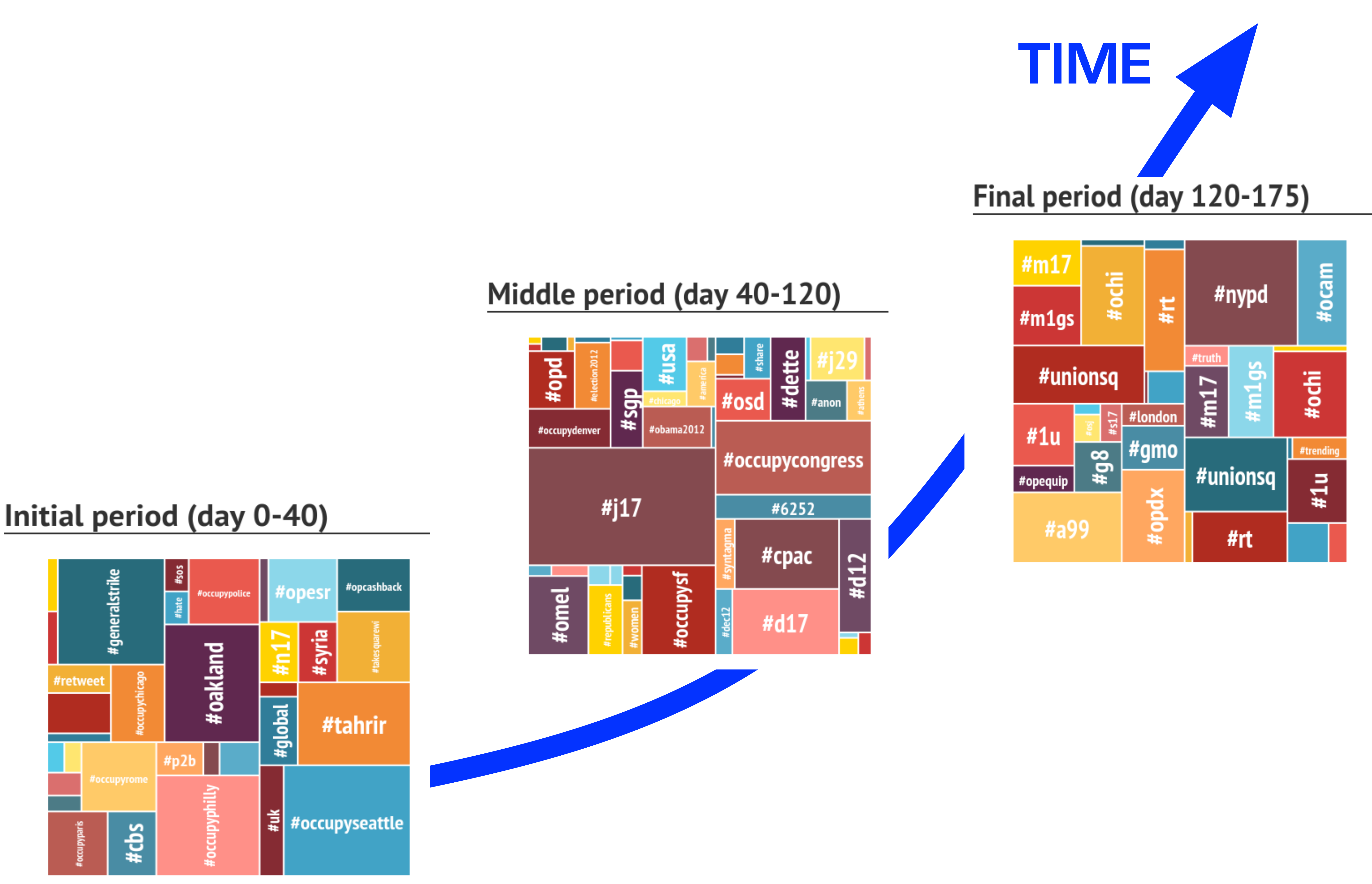}}
\caption{Top has tags in the three phases of the Occupy Wall Street Movement \label{fig8}}
\end{figure}

\begin{figure}[htpb]
\centerline{\includegraphics[width=1\textwidth]{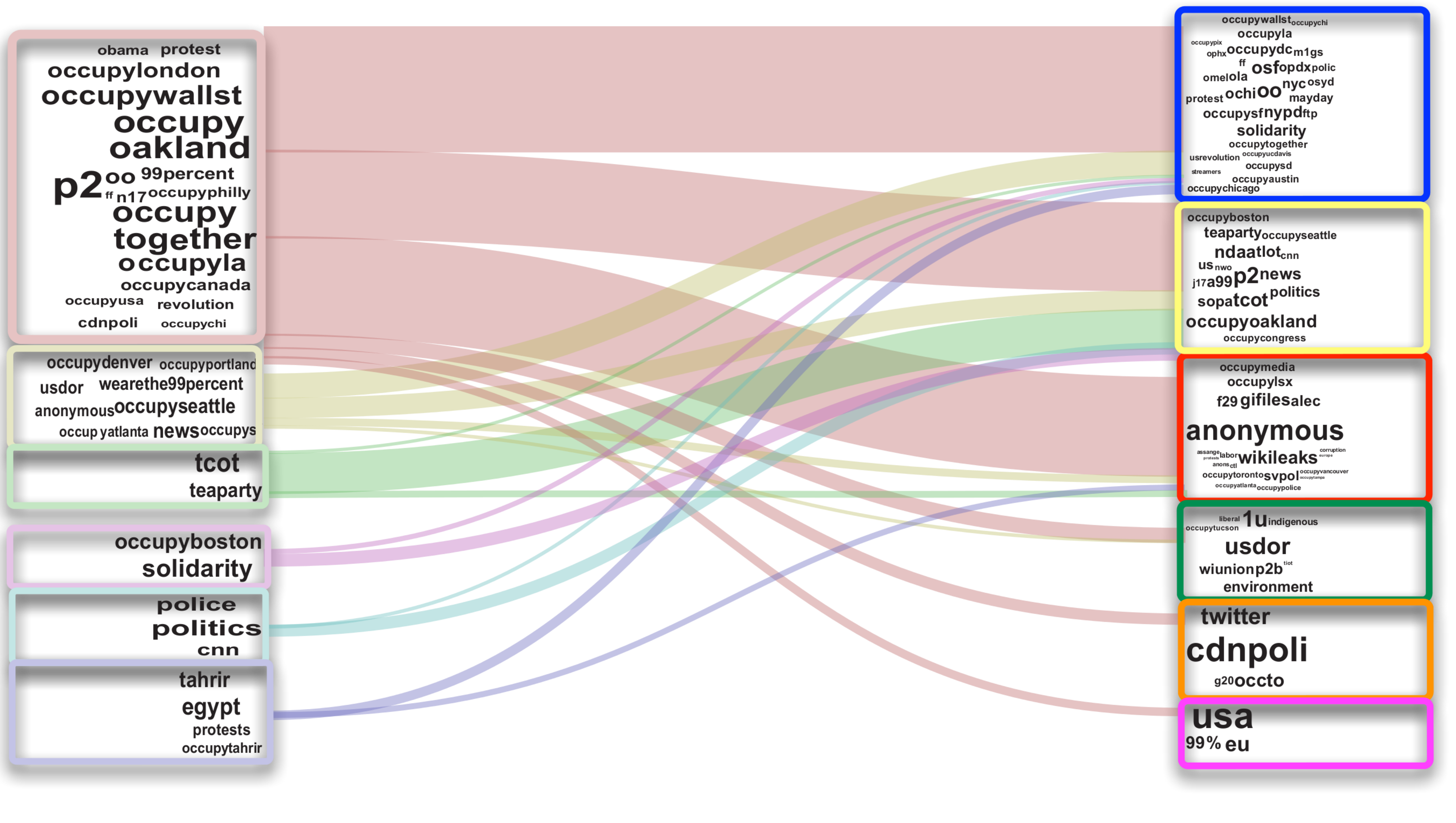}}
\caption{Evolution of the semantic communities during Occupy Wall Street Movement \label{fig9}}
\end{figure}

We also reconstructed the temporal daily networks G(V,E,t), both for the semantic structures and for the users. We studied how semantic and participatory innovation is present in the network evolution. To this end, at each time step $t$, we evaluate the difference between the network at time $t$ and $t-1$. The difference is evaluated considering the Jaccard index between the set of nodes at $t$ and at $t-1$ ($J_N(t)$) and between the set of edges at $t$ and at $t-1$ ($J_E(t)$). A low value for the index , indicates that the two networks are very dissimilar and that in between two measurements an innovation has been introduced; viceversa, a high value of the index indicates continuity in the process. Observing the trends in the temporal series of $J_N$ and $J_E$(Fig. \ref{fig7}) we can see that during the first phase of the discussion, coincident with the period of  the occupation of Zuccotti Park, the innovation level is very high (very low Jaccard index).  
both in the topics (semantic network) and in the participation (users' network). At the beginning, the Jaccard index follows a positive increasing trend and reach, after the eviction of Zuccotti Park, a stable plateau of the innovation activity. In the last phase the Jaccard index start to increase again, representing the fact that few important topics and activist remain in the discussion. \\
In Fig \ref{fig8}, we display  hashtags appearing in the initial phase
of the movement (days 1-40, around Zuccotti Park occupation) and after
disappearing, this appearing for the first time in the middle
period  (days 41-120) and then disappearing, and finally those of the last days
 (days 121-175) of the
data collection. We notice that the hashtags strictly relative to the
first phase were mostly linked to the Occupy movement
activism (\#takethesquare, \#generalstrike,\#opcashback,\#opESR) and to
local declinations of the movement that
started simultaneously to the New York movement all around the world
(\#occupyphilly,\#occupyseattle,\#occupyrome,\#occupyparis). In the middle
phase the discussion is enriched with several keywords concerning the
institutional politics (\#elections2012,\#obama,\#republicans) and with
the appointments for demonstrations (\#j17)\cite{Daniel:G9d0kFzj}. Finally, the latest phase
is characterised by several keywords concerning the political
repression of the movement and more general societal topics (notice
that the world ÓOccupyÓ never occur for the first time in this phase).
As the discussion is going on, new topics arise and old ones become less and less popular and thus users change their interests, this reflected by the use of certain topics with respect to other.
In doing so users can enjoy the discussion with others on more relevant themes and then we observe a flux of users from one community to another.  In Fig \ref{fig9}, the alluvial plot shows the flow of users among the different topics during the period under observation: the word cloud represents topics in the 6 communities at the beginning and at the end; bands repent the flows of users  form one topic to another, that is the number of users whose interests in the discussion have changed during the period. As time passes , discussion  become   more and more focused on three major theme area. Less popular topics disappear and users originally using these hashtags are moving to more mainstream topics. Besides new topics can emerge giving rise to the creation of new small communities at the end of the process

\section{Conclusions}\label{sec:Discussion}   
In this article  we have analysed a set of tweets relative to the movement "Occupy Wall Street". Using standard network tools we have built a bipartite network whose nodes are @users and \#hashtags extracted from the Twitter. We have studied the two projections to first study the robustness of the network. The analysis has shown that both projections are resilient with respect to deletion of nodes: that is under a regime of censorship , messages could still spread  among activists. 
Previous study has focused to study the social network among Twitter users , however in this case we are more interested in study  how discussion among topics has evoked. Using community detection algorithms we have found different  semantic communities, that is topics that are strongly related among them. We noticed that as the movement has moved in time, the discussion has moved form declaration of intent to media and institution reactions. The use of the Jaccard index has also allow us to establish three different phases of the discussion evolution based only on the characteristics of the networks. As discussion evolves in time users are more likely to change interests and this is reflected by the the use of hashtags in tweets. As we see at the end the discussion is focused on three main topics area. 

\section*{Acknowledgments}
This paper presents research results of the Belgian Network DYSCO
(Dynamical Systems, Control, and Optimization), funded by the
Interuniversity Attraction Poles Programme initiated by the Belgian
Science Policy Office.
Andrea Apolloni is funded by the WHO project EPIDZB44,mathematical modelling on the impact of Hepatitis B vaccination. Floriana Gargiulo  would like to thank Timoteo Carletti for useful discussions

\bibliographystyle{plain}

\bibliography{biblio2}

\end{document}